\documentclass[aps, prl, reprint, showpacs, twocolumn, 10pt, groupedaddress]{revtex4-1}

\usepackage{graphicx, amsmath, dsfont, dcolumn, bm, times, color, braket, amssymb, multirow}

\begin{document}

\bibliographystyle{apsrev4-2}

\title{Cooling of electrons via superconducting tunnel junctions and their arrays exhibiting nodal lines}

\author{Linus Aliani and Viktoriia Kornich}
\affiliation{Institute for Theoretical Physics and Astrophysics, University of W\"urzburg, 97074 W\"urzburg, Germany }

\date{\today}

\begin{abstract}
	We study theoretically a process of cooling electrons using a superconducting tunnel junction with a $\pi$ phase difference and a usual insulator or a ferroelectric in-between, and an array of such junctions with ferroelectric layers in-between. These setups have a complex structure of entropy due to nodal lines, where the density of states can be divergent or larger than for a free electron gas at a chemical potential level. We consider a small current running from the bath of electrons through the setup, where electrons have to have higher entropy, and thus remove heat from the bath. 
\end{abstract}

\maketitle

\let\oldvec\vec
\renewcommand{\vec}[1]{\ensuremath{\boldsymbol{#1}}}
{\it Introduction.--} Cooling of electrons is an essential procedure for the electronics we use every day and for the devices used in laboratories. While mK temperatures can be reached in commercial dilution refrigerators, the lower temperatures are still hard to achieve. It is especially hard to reach lower electron temperatures since dilution refrigerators cool phonons rather than electrons and additional complications are posed for dilution refridgerators in the low-gravity-environment of space missions \cite{chaudhry:cryo12}. 

The refrigerator Carnot cycle consists of two isothermal and two adiabatic processes. This thermodynamic description allows for various physical realizations of this or analogous cycles, e.g., a heat pump driven by quantum correlations \cite{holdsworth:pra22} or based on ultracold atoms \cite{roy:pre20}. As conventional electronics often becomes non-functional at extremely low temperatures, the superconducting electronics is usually used instead \cite{simbierowicz:prx24, generalov:apl24, ingla:natel25}, although, this is not a rule \cite{staveren:ieee25}. There are a number of suggestions for heat engines based on superconducting heterostructures \cite{muhonen:rpp12, nguyen:pra16, vadimov:aip22, lemziakov:jltp24, cioni:prb25}, in particular, diffusive SNS junctions \cite{rabani:prb08, vischi:scirep19} and topological Josephson junctions \cite{scharf:prr21,scharf:cp20}. Some of these works focus on phase and temperature control of the entropy and a supercurrent, some work with Andreev bound states. There are also works based on removing hot electrons from a metal via tunnelling into a superconductor \cite{giazotto:apl02, kawabata:apl13}, where one of the limiting factors is heating due to Andreev current \cite{rajauria:prl08, courtois:crp16, pimanov:bjn22, ozaeta:prb12}.

In this work, we study the opportunity for a one-time cooling of an electron bath {\it via electrons} that come from it, pass through the setup based on superconducting tunnel junctions, and leave further into the circuit, see Fig.~\ref{fig:setup} (a). Due to an applied low voltage, a small current is transferring the electrons from the bath into the circuit containing the high-entropy setup. In order to pass through the setup, the electrons need to reach the entropy of the setup, and this they do by acquiring heat from other electrons in the bath. Electrons, which transfer from the bath into the setup, therefore extract heat from the bath with themselves. Once the electrons pass the high-entropy setup, they move on through the conductors of lower entropy and therefore emit heat there. This scheme is analogous to the one described in Refs.~\cite{dolleman:abstract, dolleman:arxiv26} and allows for addressing specifically electrons instead of phonons.

\begin{figure}[b]
		\includegraphics[width=\linewidth]{setup_new.png}
		\caption{(a) The cooling scheme, where the current flows from the electron bath (which is capacitively coupled to the voltage source) through the setup with a large entropy into the electric circuit. Electrons absorb heat from the bath because they need to increase their entropy (see the dark-blue region). They emit this heat after the region with high entropy (red region). (b) Setup consisting of a tunnel junction of superconductors (SC1 and SC2) with a ferroelectric layer (FE) in-between. (c) Setup consisting of an array of tunnel junctions with ferroelectric layers in-between. }		\label{fig:setup}
\end{figure} 

We suggest to use superconducting tunnel junctions exhibiting nodal lines in order to obtain a large entropy for very low energies. We start from considering a tunnel junction consisting of two conventional superconductors with equal real mean fields and a $\pi$-phase shift, $\Delta_2=-\Delta_1=\Delta$, and an insulating layer in-between. We show that there is a point, $t=\Delta$ with $t$ being the tunnelling amplitude, where the density of states becomes infinite. Certainly, in real conditions, this peak will not be infinite, but will be restricted at least by scattering due to irregularities of the structure \cite{bruder:prb96}, e.g., in the contact regions between different materials. However, this peak would be hard to find in experiments, and therefore, we consider a tunnel junction with a ferroelectric layer in-between, see Fig.~\ref{fig:setup} (b). Ferroelectrics, when polarized, produce a spin-orbit interaction \cite{zhang:prb24}, thus the electrons' spins will rotate, while tunnelling through the ferroelectric layer. This makes the peak structure more complicated, exhibiting dependencies on chemical potential and the strength of polarization. The remnant polarization of a ferroelectric can be flipped with an external electric field \cite{izyumskaya2009oxides, pan:csr24}. It can also be varied by different means, e.g., by mechanical stress \cite{acosta:apl14} and UV light \cite{dimos1994photoinduced, warren:apl95, cao:jmc12}. 

For more fine-tuned temperature control, e.g., variation of temperature by several times, we suggest to use a multilayer structure of superconductors and ferroelectrics, with a phase difference of $\pi$ between neighbouring superconducting layers, see Fig.~\ref{fig:setup} (c). We consider both aligned and alternating polarizations of ferroelectric layers. This structure has nodal lines with a mostly linear spectrum around them, and thus an interesting dependence of the density of states (DOS) on various parameters of the system, giving different opportunities for tuning the entropy of the setup and consequently the temperature of the electron bath.    

{\it Cooling scheme.--} Here, we discuss cooling of electrons via running a current through a system featuring spatially varying entropy. We assume the currents to be tunnelling due to enough wide insulating barriers between all parts of the scheme. An electron current flows from the bath, which we want to cool down through the higher entropy region. When entering this region, i.e., at the edge between the conduction channel from the electron bath and the high-entropy setup, electrons take heat from the bath in order to increase their entropy. Taking heat away lowers the free energy of the electron bath, and consequently lowers its temperature (see the dark-blue region in Fig.~\ref{fig:setup} (a)). Thus, there is a gradient from cool (light-blue region) to even cooler temperatures (dark-blue region) in the bath. As the electron-electron scattering time is very short (can be $\sim 10^{-16}$~s, \cite{chen:pnas17}), the gradient of temperature evens much faster than the characteristic time of electrons passing through the setup, thus we consider the temperature of the bath to be the same over the whole volume in our calculation. When exiting the setup, electrons emit heat, see the red region in Fig.~\ref{fig:setup} (a).
After emitting this heat the electrons are moving further along the circuit and away from the setup. Hot electrons can therefore not flow back towards the bath. For the phonons that can be eventually heated via electrons in the red region, the path towards the bath is obstructed by the length of the setup and the phonon's comparatively short mean free path. Moreover, they can be removed by coupling the red region to a phonon bath or a conventional phonon-cooling fridge. The amount of absorbed heat can be defined from the change of energy of electrons in the bath:
\begin{eqnarray}
\label{eq:equilibrium}
Q=F(T_i)-F(T_f),
\end{eqnarray}
where $Q$ denotes the heat, $F$ is the free energy of electrons in the bath, $T_i$ and $T_f$ are the initial and final temperatures, respectively, $T_i>T_f$.
If we consider the conventional spin-degenerate two-dimensional electron gas (2DEG), the free energy is
\begin{eqnarray}
\label{eq:F(T)}
F(T)=-Tl_xl_z\int n_{\rm 2DEG}\ln{\left[1+e^{-\frac{E-\mu}{T}}\right]}dE,
\end{eqnarray}
where $l_x$ and $l_z$ are lengths in $x$ and $z$ directions, respectively, $n_{\rm 2DEG}$ is the DOS, $T$ is temperature, $\mu$ is a chemical potential, and $E$ is energy.

On the other hand, the absorbed heat is defined as
\begin{eqnarray}
\label{eq:Q}
Q=\int\left( L_xL_z T(\tau)\rho\int \Delta S(E)dE-q_{\rm loss}\right)d\tau,
\end{eqnarray}
where $\rho$ is the particle current of electrons, $\Delta S(E)$ is the change of entropy of the flowing electrons from the bath to the given setup at the energy $E$, temperature depends on the time $\tau$, and $L_x$ and $L_z$ are the lengths of the setup in $x$ and $z$ directions, respectively; $q_{\rm loss}$ denotes the heating of electrons due to various processes that happen simultaneously to the described cooling. These processes limit the lowest possible temperature.

We choose to consider 2D systems as follows from Eqs. (\ref{eq:F(T)}) and (\ref{eq:Q}), because it is more convenient for the discussion of the cooling principle, as is seen from the comparison of the densities of states below. However, the cooling principle is not confined to 2D.

The particle current of electrons $\rho$ can be defined from the electron current, $I$, as follows:
\begin{eqnarray}
\rho=\frac{I}{e}=\frac{\Sigma V}{e},
\end{eqnarray}
where $e$ is the charge of an electron, $V$ is the applied voltage, and $\Sigma$ is the conductance. As we run direct current from the bath through the setup and the outer circuit, via which the electrons escape, it should be constant at any cross-section and can be measured in experiments at the edges of the system without a disturbance of the cooling setup.

In our consideration, $q_{\rm loss}$ contains the exchange of heat with phonons \cite{wellstood:prb94},
\begin{eqnarray}
q_{\rm loss}=\mathcal{V}\Omega (T_{\rm ph}(\tau)^5-T(\tau)^5).
\end{eqnarray}
Here, $\mathcal{V}$ denotes the volume of the whole system, $\Omega$ is a constant that depends on Fermi energy of electrons and sound velocity \cite{wellstood:prb94}, and $T_{\rm ph}(\tau)=\int_0^\tau V^2\Sigma d\tau'+T_{\rm ph}(\tau=0)$ is the phonon temperature that includes the Joule heating. This Joule heating comes from all dissipative currents, which are tunnelling currents in our case due to the wide insulating barriers between the setup and the external circuit. If the opacity of the contacts is improved, we obtain transport currents with somewhat different dissipative processes, but the scheme is the same, as the heating due to the dissipation will be overcome by a very strong entropy change. 
The minimum temperature, $T_{\rm min}$, is defined by the condition, that the expression under the integral in Eq. (\ref{eq:Q}) is zero.

Eq. (\ref{eq:Q}) displays that the system contains many degrees of freedom and can be tuned in different ways. However, the most interesting effect for us comes from the change of entropy.

{\it General expression for entropy.--}
In a many-body case, the Gibbs entropy in a grand canonical ensemble is \cite{wolverton:prb94, rozen:nature21, SM},
\begin{eqnarray}
\label{eq:S}
&&S=\\ \nonumber &&-\int n(E)[f(E)\ln{(f(E))}+(1-f(E))\ln(1-f(E))] dE,
\end{eqnarray} 
where $f(E)=(1+\exp(E/T))^{-1}$ is the Fermi-Dirac distribution, $n(E)$ is the DOS of the setup. 

We consider a current running through the setup, assuming the lower potential to be applied behind it, see Fig.~\ref{fig:setup}~(a). As we need to count only the entropy of the electrons that flow, i.e., right-movers defined by the window $f(E)-f(E+eV)$, we divide Eq.~(\ref{eq:S}) by $2$ and have to integrate up to approximately $-eV$. As we consider the case of very low temperatures, the limit $eV>T$ is natural, and the integration interval will be defined by the expression in the square brackets of Eq.~(\ref{eq:S}). Thus, we simplify this term to $-\ln{2}$, but put the integration interval $\sim \pm 2T$, 
\begin{eqnarray}
\label{eq:Ssimpl}
S\simeq \ln{2}\int_0^{2T} n(E)dE,
\end{eqnarray}
where we have used the fact that $n(E)$ is symmetric with respect to $E=0$ due to the particle-hole symmetry. At very low temperatures, $T\sim 0$, the main contribution comes from the states at energies $E\simeq 0$, and $S\propto n(0)$. Thus, in case of very low temperatures, the behaviour of the entropy is defined by the DOS in the region of nodes/nodal lines.

\begin{figure}[tb]
		\includegraphics[width=0.8\linewidth]{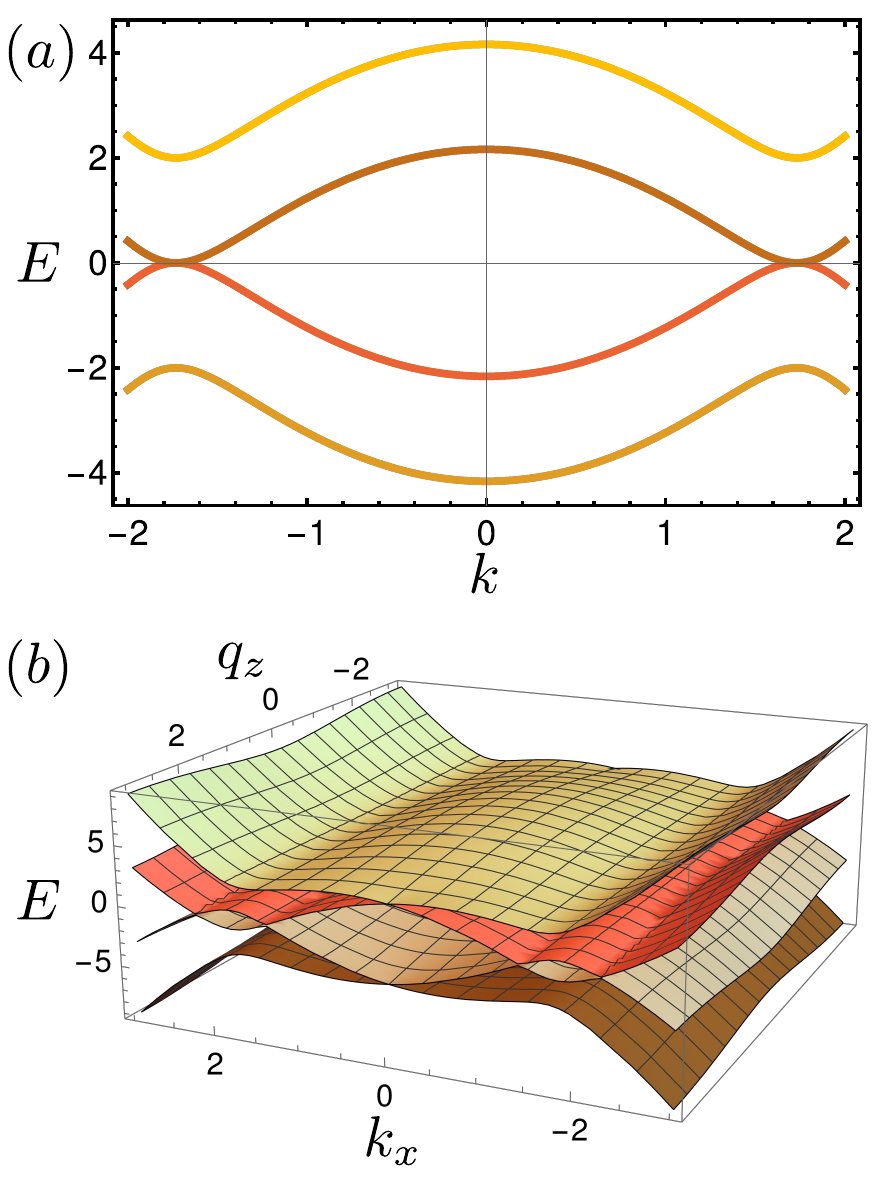}
		\caption{Spectra of (a) the Hamiltonian for a tunnel junction, $H_{\rm TJ}$, and (b) the Hamiltonian for a multilayer structure with alternating polarization of ferroelectric layers, $H_{\rm MLJ}^-$. In (a), $t=\Delta=1$, $\mu=3$, and $m=0.5$. The eigenenergies touch at $E=0$ with their spectra being quadratic around the touching point, i.e., the node. This induces a divergence in the DOS. As $k=\sqrt{k_x^2+k_y^2}$, the node is actually a nodal circle. In (b), $t_1=0.5$, $t=\sqrt{1-t_1^2}$, $\mu=3$, $\Delta=1$, $m=0.5$. The nodal lines are formed by the second and the third eigenenergies, that span through the whole range of $q_z$. For some other combinations of parameters, they start and end inside of $q_z\in [-\pi,\pi]$. The spectra around these nodal lines are usually not quadratic, except for some special points for certain parameters, therefore the DOS is not overall infinite.}		\label{fig:spectrum}
\end{figure} 
	
{\it Entropy of a superconducting tunnel junction.--} Let us consider a junction of two conventional superconductors with a thick insulating layer in-between. The thickness of the insulating layer should be enough for the junction to be within the tunnelling regime (unlike for Josephson junctions), i.e., there is no resonant scattering of electrons and holes that form Andreev bound states, but random tunnelling between superconducting layers. The Hamiltonian of such a junction in the basis $\Psi^\dagger_{\bm k}=\{c^\dagger_{1,\bm k,\uparrow},c^\dagger_{1,\bm k,\downarrow}, c_{1,-{\bm k}, \uparrow}, c_{1,-{\bm k},\downarrow}, c^\dagger_{2,\bm k,\uparrow},c^\dagger_{2,\bm k,\downarrow}, c_{2,-{\bm k}, \uparrow}, c_{2,-{\bm k},\downarrow}\}$ is,
\begin{eqnarray}
H_{\rm TJ}=\sum_{\bm k}\Psi_{\bm k}^\dagger\mathcal{H}_{\rm TJ}\Psi_{\bm k}=\sum_{\bm k}\Psi_{\bm k}^\dagger[(\frac{{\bm k}^2}{2m}-\mu)\sigma_0\tau_z\chi_0+\\ \nonumber+\frac{1}{2}[\Delta_1(\chi_0+\chi_z)+\Delta_2(\chi_0-\chi_z)]\sigma_y\tau_y^*+t\sigma_0\tau_z\chi_x]\Psi_{\bm k},
\end{eqnarray} 
 where $\bm{k}=\{k_x,k_y\}$ is the momentum of electrons in 2D, $m$ is their effective mass, $\sigma$, $\tau$, and $\chi$ denote spin, particle-hole, and layer subspaces, respectively. For shortness of notation, we take the tunnel coupling $t>0$. The superconducting mean fields $\Delta_1$ and $\Delta_2$ describe superconductivity in the layers 1 and 2, respectively, and are taken to be real.

As we are interested in energies around the chemical potential, let us consider the case, when the quasiparticle gap of such a junction can close, i.e., $\Delta_2=-\Delta_1=\Delta$. The spectrum of this system has four bands,
\begin{eqnarray}
E=\pm t\pm\sqrt{\left(\frac{k_x^2+k_y^2}{2m}-\mu\right)^2+\Delta^2},
\end{eqnarray}
where each band is doubly degenerate. We can see that the condition $E=0$ can be fulfilled for the kinetic energy equal to $\mu$ and $t=\Delta$, see Fig.~\ref{fig:spectrum}~(a). We calculate the DOS for this junction as, 
\begin{eqnarray}
n(E)=\sum_{j}\int\frac{dk_xdk_y}{(2\pi)^2}\delta(E-E(k_x,k_y,j)),
\end{eqnarray} 
 where $j=\{1,..,8\}$. For $E=0$, $t> \Delta$ and $\mu\geq \sqrt{t^2-\Delta^2}$, we obtain
 \begin{eqnarray}
 \label{eq:DOSTJ}
 n(0)=\frac{2t}{\pi\sqrt{t^2-\Delta^2}}.
 \end{eqnarray} 
 For $\mu<\sqrt{t^2-\Delta^2}$ and $t>\Delta$, the DOS is halved, but this exotic limit is not considered here. For $t<\Delta$, there are no crossings of the eigenenergies and $E=0$, and thus $n(0)$ is zero.
 
From Eq.~(\ref{eq:DOSTJ}) follows that there is a divergence at $t\rightarrow\Delta$. It is a consequence of the eigenenergies touching $E=0$ by their minima, see Fig.~\ref{fig:spectrum} (a). Note that the touching points have a quadratic spectrum around them, giving an infinite DOS instead of zero, as for usual nodal superconductors with linear spectra around the nodes \cite{won:aip05, alrub:physb08}. The infinite DOS implies a strongly increased entropy and that this setup can be used for an extremely effective cooling of electrons, once this zero-point is found. However, tuning $\Delta$ or $t$ is not trivial in an experiment. Therefore we explore other setups motivated by this one.

{\it Entropy of a superconducting tunnelling junction with a ferroelectric in-between.--} Ferroelectrics are usually insulators \cite{wemple1970polarization, piskunov2004bulk}, therefore internal electric polarization does not induce current. 
This polarization can induce Rashba spin-orbit interaction \cite{zhang:prb24} similar to electric fields in triangular quantum wells. The electric polarization acts on the tunnelling electrons, even though the electrons pass through the insulating gap. Due to their high velocity, spin-orbit interaction is not negligibly small, and their spins rotate. This means, that we have not only spin-diagonal terms $t$ for tunnelling, but also off-diagonal terms, $t_1$, proportional to the Rashba spin-orbit term. Then the Hamiltonian is
\begin{eqnarray}
H_{\rm JFE}=\sum_{\bm k}\Psi_{\bm k}[\mathcal{H}_{\rm TJ}+t_1(k_y\sigma_x\tau_0-k_x\sigma_y\tau_z)\chi_x]\Psi_{\bm k}.\ \ \
\end{eqnarray}
This Hamiltonian has a non-degenerate spectrum, and thus more distinct zero touching points of the eigenenergies. This means, that the DOS will have additional peaks and dependence on other parameters of the system, e.g., the chemical potential $\mu$ that can be varied by applying an electrostatic potential, the strength of polarization that affects $t_1$ directly and can be changed by external means. In Fig.~(\ref{fig:DOS}) (blue lines), we show how the DOS of this junction depends on $\mu$, $t_1$, and $\Delta$. Thus, this setup allows for more operational phase space, while still exhibiting divergences in the DOS.

However, what if we still have issues tuning parameters into those peaks or we do not need a strong cooling effect, but cooling by a fraction of the current temperature? For this purpose, we consider a more complicated setup, consisting of a large number of superconducting and ferroelectric layers.

\begin{figure}[tb]
		\includegraphics[width=0.75\linewidth]{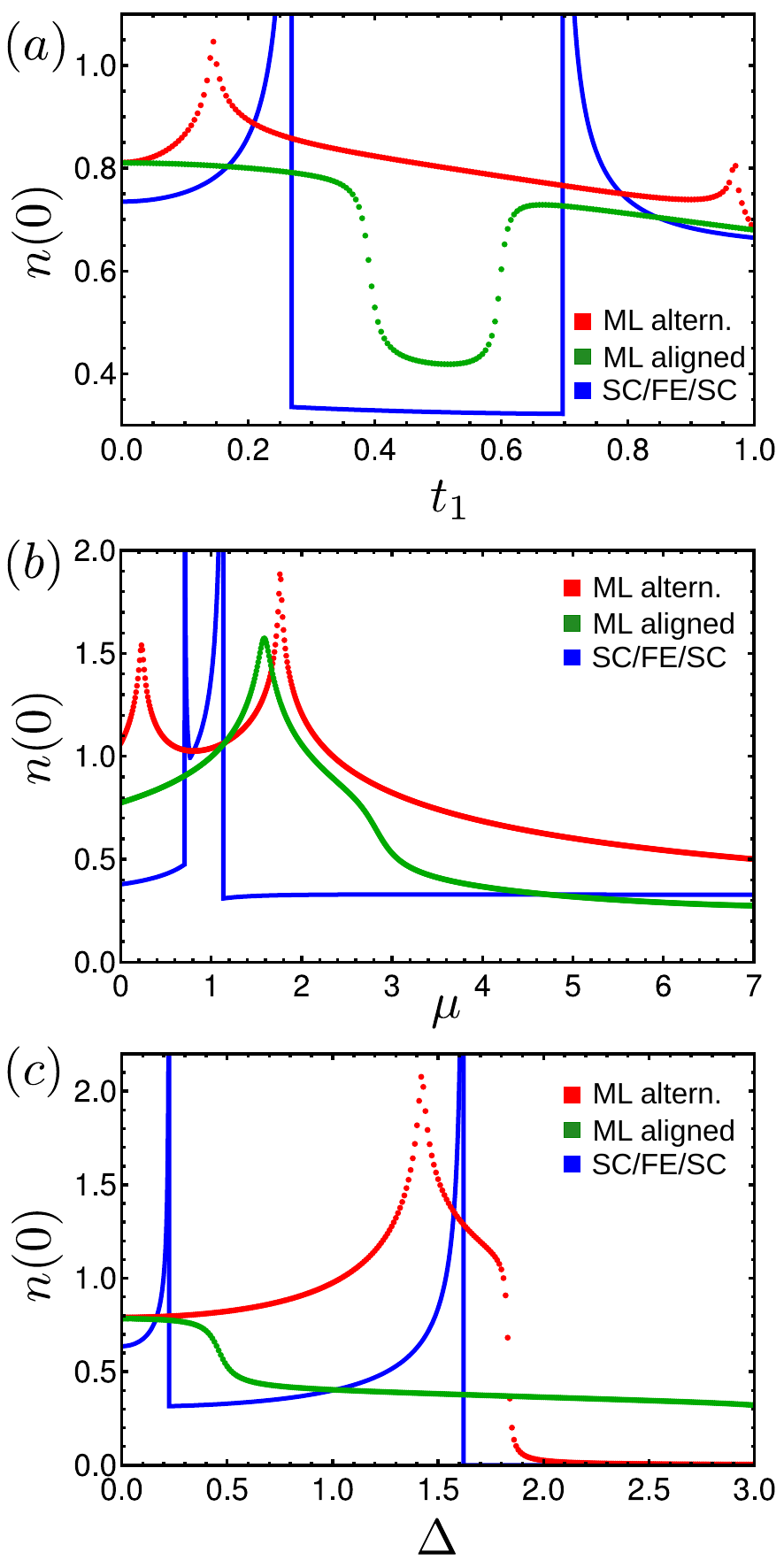}
		\caption{Density of states of a tunnel junction superconductor-ferroelectric-superconductor (SC/FE/SC, blue lines) with the phase difference $\pi$, and of a multilayer structure of superconducting layers with alternating phases $0$ and $\pi$ and ferroelectric layers with aligned polarization (green dots, $\eta=0.05$), and with alternating polarization (red dots, $\eta=0.01$). In all subplots, $t=\sqrt{1-t_1^2}$, $m=0.5$. We use (a) $\mu=3$, $\Delta=0.5$; (b) $t_1=0.4$, $\Delta=0.5$; (c) $\mu=3$, $t_1=0.4$. In (a), a multilayer structure with alternating polarization gives the most stable values overall. In (b), the behavior of the dependences on $\mu$ is very different for these parameters, but all the curves flatten for larger $\mu$. In (c), the behavior is also very different, the multilayer structure with aligned polarization does not give zero for large $\Delta$ in contrast to the other two setups. We note that the results differ a lot for different parameters, including the amount of peaks/drops, because the spectra change.}		
		\label{fig:DOS}
\end{figure}

{\it Multilayer structure of superconductors and ferroelectrics.-- } Here, we consider thin quasi-1D superconducting slabs and layers of ferroelectric in-between, that are thick enough to ensure tunnelling regime \cite{kornich:prr19, kornich:prb20}, see Fig.~\ref{fig:setup}~(c). In the following, we will omit $k_y$ and $k_z$ of electrons in the superconducting slabs, because they are small and almost constant in a quasi-1D system. In the kinetic energy, they can be absorbed into $\mu$. Due to the confinement in the $y$ dimension, electron velocity is too small for a sizeable spin-orbit interaction with $k_y$. 

Let us consider this layered structure infinite in $z$ dimension, analogously to infinite multilayer systems discussed in Refs.~\cite{burkov:prl11, meng:prb17}, i.e., it works as a crystal with properties defined by the combination of properties of contributing layers. We have two obvious options: identical polarization of all ferroelectric layers and alternating polarization. The Hamiltonian of such a multilayer structure is \cite{SM},
\begin{eqnarray}
H_{\rm MLJ}^\pm=\sum_{k_x,q_z}\Psi_{k_x,q_z}^\dagger \left[\left(\frac{{k_x}^2}{2m}-\mu\right)\sigma_0\tau_z\chi_0+\right.\\ \nonumber+\frac{1}{2}[\Delta_1(\chi_0+\chi_z)+\Delta_2(\chi_0-\chi_z)]\sigma_y\tau_y^*+\\ \nonumber+t\sigma_0\tau_z[(1+e^{-iq_z})\chi_++(1+e^{iq_z})\chi_-]+\\ \nonumber+t_1k_x\sigma_y\tau_z[(1\pm e^{-iq_z})\chi_++(1\pm e^{iq_z})\chi_-]\Psi_{k_x,q_z},
\end{eqnarray}
where $q_z\in [-\pi,\pi]$ is the wave vector accounting for the periodicity of our structure, and $\pm$ stand for the aligned and alternating polarizations, respectively. The alternating polarization gives a minus sign, because the Hamiltonian for spin-orbit interaction changes its global sign with the change of the sign of the electric field that induces it. In this case, the spectrum is twice degenerate, because electrons in each superconducting layer experience local space inversion symmetry in $z$ direction.

The spectrum of $H_{\rm MLJ}$ has nodal lines, see Fig.~\ref{fig:spectrum} (b) for the spectrum of $H_{\rm MLJ}^-$. We can see that the spectrum around these nodal lines is mainly not smooth, because the eigenenergies are defined via the absolute values,
\begin{eqnarray}
E_{\rm MLJ-}=\pm\Bigg|\sqrt{\left(\frac{k_x^2}{2m}-\mu\right)^2+\Delta^2}-\\ \nonumber-\sqrt{2(t^2+k_x^2t_1^2+(t^2-k_x^2t_1^2)\cos{q_z})}\Bigg|,
\end{eqnarray}
and the derivatives of eigenenergies with respect to $k_x$ and $q_z$ in the majority of cases do not exist strictly in the nodes. The DOS in such cases is considered to be zero \cite{won:aip05, alrub:physb08}. However, a multilayer structure will most likely have numerous irregularities in the contact regions, and the scattering due to them expands nodal lines into areas. In order to take this into account we calculate the DOS via the Green's function,
\begin{eqnarray}
n(0)=-\frac{1}{\pi}\Im[{\rm Tr}(G(0))],
\end{eqnarray}
where $G(E)=(E-H_{\rm MLJ}+i\eta)^{-1}$, and $\eta$ is a small finite number representing the presence of irregularities in the multilayer structure. 

We calculate the DOS for aligned and alternating polarizations, see Fig.~\ref{fig:DOS} green and red dots, respectively. We can see that the alternating polarization gives a rather stable and large DOS in the dependence on $t_1$, Fig.~\ref{fig:DOS}~(a). It is rather stable again and larger than the DOS for the other setups in the dependence on $\mu$, Fig.~\ref{fig:DOS}~(b). In the dependence on $\Delta$, Fig.~\ref{fig:DOS}~(c), the aligned polarization gives a smoother DOS, than the other setups, but it is also overall smaller. For comparison, the DOS of a spin-degenerate non-interacting 2DEG, is $n_{\rm 2DEG}=1/(2\pi)$ without taking into account particle-hole doubling. With the doubling, $n_{\rm 2DEG}=1/\pi$ for $0\leq E\leq \mu$ and $1/(2\pi)$ for $0\leq \mu<E$. This variety of regimes can be effectively used for fine-tuning of temperature, especially as the polarization of ferroelectrics can be flipped by an external electric field. Thus, the setups require a very limited number of material specific assumptions. This provides a promising platform for cooling of electrons for a wide variety of material-specific and external conditions.

{\it Estimation of an operating time.--} From the discussion above follows that, if $n(E)\rightarrow \infty$, we will obtain a desired low temperature very quickly. However, junctions are not ideal, therefore the peaks are not infinite, and we need to run the current for a certain time in order to reach this temperature. Here, we will make an analytical evaluation of this time. For that, we first represent Eq.~(\ref{eq:Ssimpl}) in terms of a finite difference, because $T$ is very small,
\begin{eqnarray}
S=\ln{2}[n(0)+n(2T)]T,
\end{eqnarray}
then, taking into account that $n(E)$ changes smoothly and the special points, i.e., divergences, are cut due to imperfections of the structure, we take $n(0)\simeq n(2T)$, substitute it into Eq.~(\ref{eq:Q}) obtaining,
\begin{eqnarray}
&&Q\simeq\\ \nonumber&&\int_0^{\Delta \tau} \left[\frac{2\ln{2}L_xL_z\Sigma V}{e} (n(0)-n_{\rm 2DEG})T^2(\tau)-q_{\rm loss}\right]d\tau.
\end{eqnarray}
As we run a small current through the system, temperature will not change abruptly, therefore we can express the last integral as a finite difference too. We also neglect $T^5$ compared to $T_{\rm ph}^5$ in $q_{\rm loss}$. This allows for the estimation of the operating time, $\Delta \tau$:
\begin{eqnarray}
\Delta \tau\simeq\frac{e(F(T_i)-F(T_f))}{\ln{2}L_xL_z\Sigma V(n(0)-n_{\rm 2DEG})(T_i^2+T_f^2)-q_{\rm loss}},
\end{eqnarray} 
where $T_f>T_{\rm min}$.

{\it Conclusions.--} In this work, we have described the scheme of cooling electrons to extremely low temperatures using the setups of superconducting tunnel junctions and their arrays with a $\pi$ phase difference between the neighbouring superconducting layers. We employ the obtained nodal lines and the states around them for running current from the electron bath, which we want to cool down, through this setup to the outside. The increase in entropy in the setup forces electrons to absorb heat from the bath. Notably, a conventional superconducting tunnel junction gives an infinite density of states at $\Delta=t$, and thus a strongly increased entropy. The tunnel junctions with ferroelectric in-between or the arrays of such junctions provide us a larger phase space and a variety of working regimes, that can be changed externally. The main principle of the presented cooling scheme is based on the entropy difference, and thus can be tailored to a wide variety of materials, including conventional and unconventional superconductors, different ferroelectrics, and outer circuit materials.

\begin{acknowledgments}
	We acknowledge useful discussions with Shun Tamura, Robin Dolleman, Laurens Molenkamp, Bj\"orn Trauzettel, and Vasilii Vadimov. This work was supported by the Würzburg-Dresden Cluster of Excellence ct.qmat, EXC2147, project-id 390858490, and the DFG (SFB 1170). \end{acknowledgments}

	\begin{widetext}
\section*{Supplemental Material }

\maketitle
\renewcommand{\theequation}{S\arabic{equation}}
\setcounter{equation}{0}
\renewcommand{\thefigure}{S\arabic{figure}}
\renewcommand{\figurename}{Supplementary Fig.}

\setcounter{figure}{0}
\renewcommand{\thesection}{S\arabic{section}}
\setcounter{section}{0}

In this Supplemental Material, we present additional details and calculations regarding: 1) derivation of Gibbs entropy for a system of fermions; 2) derivation of the Hamiltonian for the multilayer system.

\section{S1. Gibbs entropy for a system of fermions}
In this section, we will derive the expression for Gibbs entropy for a system with many fermions, which obey a quadratic Hamiltonian. It can be, e.g., free electron gas or a BCS-type superconductor. By definition, free energy and entropy are defined as follows:
\begin{eqnarray}
\label{Seq:Entropy1}
	F &= &-\frac{1}{\beta}\ln(Z),\\
\label{Seq:Entropy2}
	S &= &-\frac{\partial F}{\partial T} = -\frac{\partial \beta}{\partial T}\frac{\partial F}{\partial \beta} = \beta^2\frac{\partial F}{\partial \beta} =  \beta^2\left( \frac{1}{\beta^2} \ln(Z) - \frac{1}{\beta}\frac{\partial \ln(Z)}{\partial \beta} \right),
\end{eqnarray}
where $\beta$ is an inverse temperature, $\beta=1/T$.
The quadratic Hamiltonian is
\begin{equation}
\begin{aligned}
\label{Seq:H}
	\hat{H} = & \sum_{i}^{N} (\varepsilon_i - \mu)a_i^{\dagger}a_i,
\end{aligned}
\end{equation}	
where $i$ denotes eigenstates, $N$ is an overall number of eigenstates, $\varepsilon_i$ is the eigenenergy of the $i$th eigenstate, $a_i$ are the fermionic operators, $\mu$ is the chemical potential. Then the partition function is,
\begin{equation}
	\begin{aligned}
	Z = &\sum_{\{\Phi\}}\bra{\Phi}e^{-\beta \hat{H}}\ket{\Phi} = \sum_{n_i}e^{-\beta\sum_{i}^{N} (\varepsilon_i - \mu)n_i} = \sum_{n_i}\prod_{i}^{N}e^{-\beta(\varepsilon_i - \mu)n_i}\\
	=& \sum_{n_1}e^{-\beta(\varepsilon_i - \mu)n_1}\cdot\sum_{n_2}e^{-\beta(\varepsilon_i - \mu)n_2}\cdot\sum_{n_3}e^{-\beta(\varepsilon_i - \mu)n_3}...\\
	=&\prod_{i}^{N}(1+e^{-\beta(\varepsilon_i - \mu)}),
\end{aligned}
\end{equation}
where $\ket{\Phi}$ are the many-particle eigenstates of the Hamiltonian from Eq.~(\ref{Seq:H}) and $n_i \in \{0,1\}$ is the occupation number of the $i$th eigenstate. Now we can express the terms from Eq.~(\ref{Seq:Entropy2}) as
\begin{eqnarray}
\label{Seq:lnZD1}
	\ln(Z)& = &\sum_{i}^{N}\ln\left( 1+e^{-\beta(\varepsilon_i - \mu)} \right),\\
\label{Seq:lnZD2}
	\frac{\partial \ln(Z)}{\partial \beta} &= &\frac{\partial }{\partial \beta}\sum_{i}^{N}\ln\left( 1+e^{-\beta(\varepsilon_i - \mu)} \right) = \sum_{i}\frac{-(\varepsilon_i - \mu)e^{-\beta(\varepsilon_i - \mu)}}{\left( 1+e^{-\beta(\varepsilon_i - \mu)} \right)} = \frac{1}{\beta}\sum_{i}\ln(e^{-\beta(\varepsilon_i - \mu)})n_F(\varepsilon_i).
\end{eqnarray}
We also employ the following relations:
\begin{eqnarray}
(1 + e^{-x}) &=& \left(\frac{1}{1 + e^{-x}}\right)^{-1} = \left(\frac{1}{1 + e^{-x}}-1+1\right)^{-1} = \left(\frac{1 - (1 + e^{-x})}{1 + e^{-x}}+1\right)^{-1} = (-n_F(x)+1)^{-1},\\
e^{-x} &=& e^{-x} +1-1 = (e^{-x} +1)-1 = (-n_F(x)+1)^{-1} -1 = \frac{1-(1-n_F(x))}{1-n_F(x)} = \frac{n_F(x)}{1-n_F(x)}.
\end{eqnarray}

Substituting all the above into Eqs.~(\ref{Seq:lnZD1}), (\ref{Seq:lnZD2}) gives,
\begin{eqnarray}
	\ln(Z) & = &\sum_{i}\ln\left( 1+e^{-\beta(\varepsilon_i - \mu)} \right) = -\sum_{i}\ln\left( 1-n_F(\varepsilon_i) \right) \\
	\frac{\partial \ln(Z)}{\partial \beta} & = & \frac{1}{\beta}\sum_{i}\ln(e^{-\beta(\varepsilon_i - \mu)})n_F(\varepsilon_i) = \frac{1}{\beta}\sum_{i}\ln(\frac{n_F(x)}{1-n_F(x)})n_F(x).
\end{eqnarray}

Once we substitute these results into (\ref{Seq:Entropy2}):
\begin{eqnarray}
	S  & = & \beta^2\left( \frac{1}{\beta^2} \ln(Z) - \frac{1}{\beta} \frac{1}{Z}\frac{\partial Z}{\partial \beta} \right) = -\sum_{i}\ln\left( 1-n_F(\varepsilon_i) \right) -  \sum_{i}\ln(\frac{n_F(x)}{1-n_F(x)})n_F(x)=\\
	& =&-\sum_{i}^{N} [(1-n_F(\varepsilon_i))\ln(1-n_F(\varepsilon_i))+n_F(\varepsilon_i)\ln(n_F(\varepsilon_i))].
\end{eqnarray}

\section{S2. Hamiltonian for the multilayer system}
We divide the multilayer system into elementary unit cells, comprising two superconductors and one ferroelectric layer, as shown in Fig.~\ref{Sfig:setup}.
  \begin{figure}[htbp]
    \begin{center}
      \includegraphics[width=0.3\linewidth]{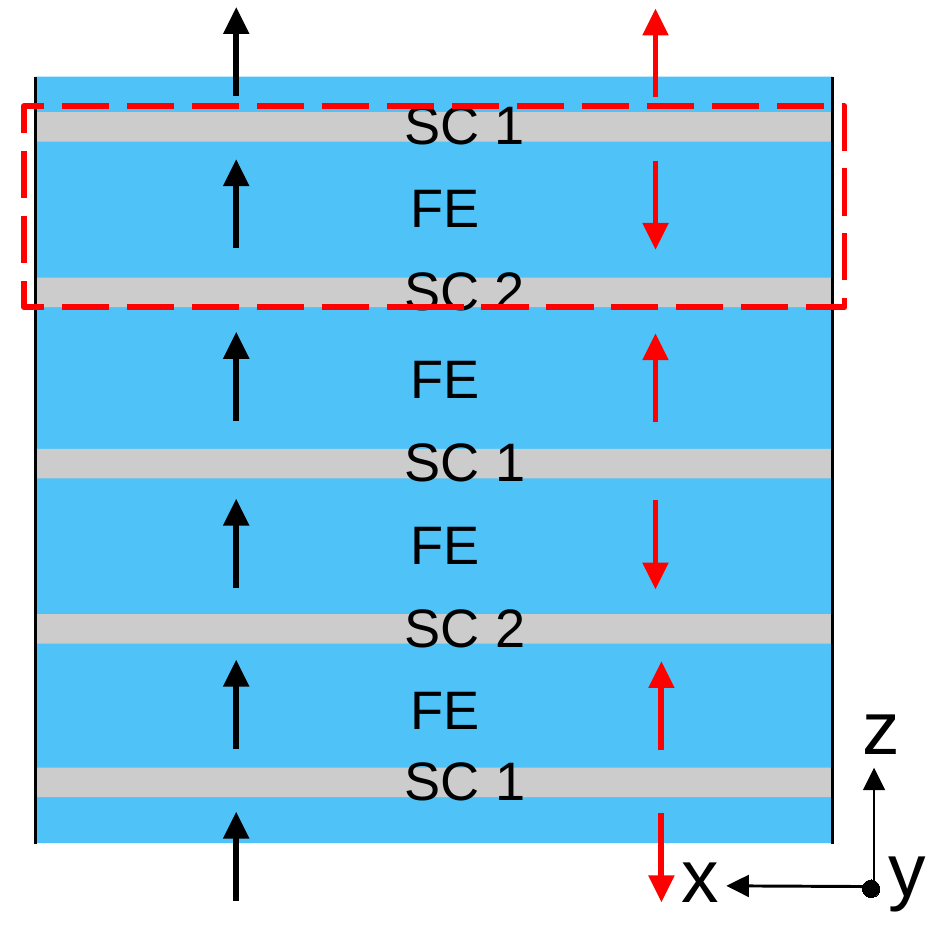}
    \end{center}
    \caption{Multilayer structure consisting of superconductors of two kinds (SC1 and SC2) and ferroelectric layers (FE), which can have an aligned polarization (black arrows) or an alternating one (red arrows). The red dashed box shows the unit cell of this structure.}
    \label{Sfig:setup}
  \end{figure}
These unit cells are connected by the layers of ferroelectric that are described by the same tunnelling Hamiltonian in case of aligned polarization and by the Hamiltonian with the global minus sign for the alternating polarization. Taking into account that there are SC1 and SC2, we obtain a tunnelling Hamiltonian similar to SSH model that gives,
\begin{eqnarray}
H_t=\sum_jt[a_{1,j}^\dagger a_{2,j}+a_{2,j}^\dagger a_{1,j+1}+H.c.],
\end{eqnarray}
where $j$ denotes the unit cells. If we perform Fourier transformation as $a_{1(2),j}=\frac{1}{\sqrt{N}}\sum_{q_z}e^{iq_zr_j}a_{1(2),q_z}$, we obtain
\begin{eqnarray}
H_t=\sum_{q_z}t[a^\dagger_{1,q_z}a_{2,q_z}+e^{iq_z}a^\dagger_{2,q_z}a_{1,q_z}+H.c.]. 
\end{eqnarray}
Thus, the tunnelling term is $t(1+e^{iq_z})$. This derivation is valid for the tunnelling terms diagonal in spin space with the prefactor $t$, and for the off-diagonal in spin space tunnelling terms with the prefactor $t_1$. If the polarization is alternating, then the ferroelectric layers outside of the unit cells give the prefactors $-t_1$, i.e., the tunnelling terms $\propto t_1(1-e^{iq_z})$.

\end{widetext}


\begin{thebibliography}{}
	\bibitem{chaudhry:cryo12} {G. Chaudhry, A. Volpe, P. Camus, S. Triqueneaux, and G. Vermeulen, ``A closed-cycle dilution refrigerator for space applications'', Cryogenics {\bf 52}, 471 (2012).}
\bibitem{holdsworth:pra22}{T. Holdsworth and R. Kawai, ``Heat pump driven entirely by quantum correlation'', Phys. Rev. A {\bf 106}, 062604 (2022).}
\bibitem{roy:pre20}{A. Roy and A. Eckardt, ``Design and characterization of a quantum heat pump in a driven quantum gas'', Phys. Rev. E {\bf 101}, 042109 (2020).}
\bibitem{simbierowicz:prx24}{S. Simbierowicz, M. Borrelli, V. Monarkha, R. E. Lake, ``Inherent Thermal-Noise Problem in Addressing Qubits'', PRX Quantum {\bf 5}, 030302 (2024). }
\bibitem{generalov:apl24}{A. A. Generalov, K. L. Viisanen, J. Senior, B. R. Ferreira, J. Ma, M. M\"ott\"onen, M. Prunnila, H. Bohuslavskyi, ``Wafer-scale CMOS-compatible graphene Josephson field-effect transistors'', Appl. Phys. Lett. {\bf 125}, 012602 (2024). }
\bibitem{ingla:natel25}{J. Ingla-Ayn\'es, Y. Hou, S. Wang, E.-D. Chu, O. A. Mukhanov, P. Wei, and J. S. Moodera, ``Efficient superconducting diodes and rectifiers for quantum circuitry'', Nat. Electr. {\bf 8}, 411 (2025).}
\bibitem{staveren:ieee25}{J. van Staveren, L. Enthoven, P. L. Bavdaz, M. Meyer, C. D\'eprez, V. Nuutinen, R. Lake, D. Degli Esposti, C. Carlsson, A. Tosato, J. Gong, B. Prabowo, M. Babaie, C. G. Almudever, M. Veldhorst, G. Scappucci, and F. Sebastiano, ``Cryo-CMOS Bias-Voltage Generation and Demultiplexing at mK Temperatures for Large-Scale Arrays of Quantum Devices'', IEEE Trans. Quantum Eng. {\bf 6}, 1 (2025).}
\bibitem{muhonen:rpp12}{J. T. Muhonen, M. Meschke, and J. P. Pekola, ``Micrometre-scale refrigerators'', Rep. Prog. Phys. {\bf 75}, 046501 (2012).}
\bibitem{nguyen:pra16}{H. Q. Nguyen, J. T. Peltonen, M. Meschke, and J. P. Pekola, ``Cascade Electronic Refrigerator Using Superconducting Tunnel Junctions'', Phys. Rev. Applied {\bf 6}, 054011 (2016).}
\bibitem{vadimov:aip22}{V. Vadimov, A. Viitanen, T. M\"orstedt, T. Ala-Nissila, M. M\"ott\"onen, ``Single-junction quantum-circuit refrigerator'', AIP Adv. {\bf 12}, 075005 (2022).}
\bibitem{lemziakov:jltp24}{S. A. Lemziakov, B. Karimi, S. Nakamura, D. S. Lvov, R. Upadhyay, C. D. Satrya, Z.‑Y. Chen, D. Subero,Y.‑C. Chang, L. B. Wang, and J. P. Pekola, ``Applications of Superconductor–Normal Metal Interfaces'', J Low Temp. Phys. {\bf 217}, 54 (2024).}
\bibitem{cioni:prb25}{F. Cioni and F. Taddei, ``High-performance Andreev interferometer based electronic coolers'', Phys. Rev. B {\bf 112}, 035402 (2025).}
\bibitem{rabani:prb08}{H. Rabani, F. Taddei, O. Bourgeois, R. Fazio, and F. Giazotto, ``Phase-dependent electronic specific heat of mesoscopic Josephson junctions'', Phys. Rev. B {\bf 78}, 012503 (2008).} 
\bibitem{vischi:scirep19}{F. Vischi, M. Carrega, P. Virtanen, E. Strambini, A. Braggio, and F. Giazotto, ``Thermodynamic cycles in Josephson junctions'', Sci. Rep. {\bf 9}, 3238 (2019).}
\bibitem{scharf:cp20}{B. Scharf, A. Braggio, E. Strambini, F. Giazotto, and E. M. Hankiewicz, ``Topological Josephson heat engine'', Commun. Phys. {\bf 3}, 198 (2020). }
\bibitem{scharf:prr21}{B. Scharf, A. Braggio, E. Strambini, F. Giazotto, and E. M. Hankiewicz, ``Thermodynamics in topological Josephson junctions'', Phys. Rev. Research {\bf 3}, 033062 (2021). }
\bibitem{giazotto:apl02}{F. Giazotto, F. Taddei, R. Fazio, and F. Beltram, ``Ultraefficient cooling in ferromagnet–superconductor microrefrigerators'', Appl. Phys. Lett. {\bf 80}, 3784 (2002).}
\bibitem{kawabata:apl13}{S. Kawabata, A. Ozaeta, A. S. Vasenko, F. W. J. Hekking, and F. S. Bergeret, ``Efficient electron refrigeration using superconductor/spin-filter devices'', Appl. Phys. Lett. {\bf 103}, 032602 (2013).}
\bibitem{rajauria:prl08}{S. Rajauria, P. Gandit, T. Fournier, F. W. J. Hekking, B. Pannetier, and H. Courtois, ``Andreev Current-Induced Dissipation in a Hybrid Superconducting Tunnel Junction'', Phys. Rev. Lett. {\bf 100}, 207002 (2008).} 
\bibitem{courtois:crp16}{H. Courtois, H. Q. Nguyen, C. Winkelmann, and J. P. Pekola, ``High-performance electronic cooling with superconducting
tunnel junctions'', C. R. Physique, {\bf 17}, 1139 (2016).} 
\bibitem{pimanov:bjn22}{D. A. Pimanov, V. A. Frost, A. V. Blagodatkin, A. V. Gordeeva, A. L. Pankratov, and L. S. Kuzmin, ``Efficiency of electron cooling in cold-electron bolometers with traps'', Beilstein J. Nanotechnol. {\bf 13}, 896 (2022).} 
\bibitem{ozaeta:prb12}{A. Ozaeta, A. S. Vasenko, F. W. J. Hekking, and F. S. Bergeret, ``Electron cooling in diffusive normal metal–superconductor tunnel junctions with a spin-valve
ferromagnetic interlayer'', Phys. Rev. B {\bf 85}, 174518 (2012).}
\bibitem{dolleman:abstract}{R. J. Dolleman, A. Rothstein, A. Fischer, L. Klebl, K. Watanabe, T. Taniguchi, D. M. Kennes, F. Libisch, B. Beschoten, C. Stampfer, ``Towards sub-mK Electron Temperatures Using Pomeranchuk Cooling in Twisted Bilayer Graphene'', 36th International Conference on the Physics of Semiconductors, (2024).}
\bibitem{dolleman:arxiv26}{R. J. Dolleman, A. Fischer, L. Klebl, A. Rothstein, D. M. Kennes, B. Beschoten, F. Libisch, C. Stampfer, ``Feasibility study of continuous electronic Pomeranchuk cooling with a flavor-degenerate Wigner crystal'', arXiv:2605.31307 (2026).}
\bibitem{bruder:prb96}{C. Bruder, W. Belzig and G. Sch\"on, ”Local Density of States in a Dirty Normal Metal Connected to a Superconductor”, Phys. Rev. B {\bf 54}, 9443 (1996).}
\bibitem{zhang:prb24}{X. Zhang, P. Zhao, and F. Liu, ``Ferroelectric topological superconductor: $\alpha$-In$_2$Se$_3$'', Phys. Rev. B {\bf 109}, 125130 (2024).}
\bibitem{pan:csr24}{Q. Pan, Z.-X. Gu, R.-J. Zhou, Z.-J. Feng, Y.-A. Xiong, T.-T. Sha, Y.-M. You, and R.-G. Xiong, ``The past 10 years of molecular ferroelectrics: structures, design, and properties'', Chem. Soc. Rev. {\bf 53}, 5781 (2024).}
\bibitem{izyumskaya2009oxides}{N. Izyumskaya, Y. Alivov and H.  Morko{\c{c}}, "Oxides, oxides, and more oxides: high-$\kappa$ oxides, ferroelectrics, ferromagnetics, and multiferroics", Crit. Rev. Solid State Mater. Sci. {\bf 34}, 89 (2009).}
\bibitem{acosta:apl14}{A. G. Acosta, J. A. Rodriguez, T. Nishida, ``Mechanical stress effects on Pb(Zr,Ti)O$_3$ thin-film ferroelectric capacitors embedded in a standard complementary metal-oxide-semiconductor process'', Appl. Phys. Lett. {\bf 104}, 222908 (2014).}
\bibitem{dimos1994photoinduced}{D. Dimos, W. L. Warren, M. B. Sinclair, B. A. Tuttle and R. W. Schwartz "Photoinduced hysteresis changes and optical storage in (Pb, La)(Zr, Ti)O$_3$ thin films and ceramics", J. Appl. Phys. {\bf 76}, 4305 (1994).}
\bibitem{warren:apl95}{W. L. Warren, D. Dimos, G. E. Pike, B. A. Tuttle, M. V. Raymond, R. Ramesh, and J. T. Evans Jr., ``Voltage shifts and imprint in ferroelectric capacitors'', Appl. Phys. Lett. {\bf 67}, 866 (1995).}
\bibitem{cao:jmc12}{D. Cao, C. Wang, F. Zheng, L. Fang, W. Dong, and M. Shen, ``Understanding the nature of remnant polarization enhancement, coercive voltage offset and time-dependent photocurrent in ferroelectric films irradiated by ultraviolet light'', J. Mater. Chem. {\bf 22}, 12592 (2012).}
\bibitem{chen:pnas17}{C. Chen, Z. Tao, A. Carr, P. Matyba, T. Szilv\'asi, S. Emmerich, M. Piecuch, M. Keller, D. Zusin, S. Eich, M. Rollinger, W. You, S. Mathias, U. Thumm, M. Mavrikakis, M. Aeschlimann, P. M. Oppeneer, H. Kapteyn, and M. Murnane, ``Distinguishing attosecond electron-electron scattering and screening in transition metals'', Proc. Nat. Acad. Sci., {\bf 114}, E5300 (2017).}
\bibitem{wellstood:prb94}{F. C. Wellstood, C. Urbina, and J. Clarke, ``Hot-electron effects in metals'', Phys. Rev. B {\bf 49}, 5942 (1994).}
\bibitem{wolverton:prb94}{C. Wolverton and A. Zunger, ``First-principles theory of short-range order, electronic excitations, and spin polarization in Ni-V and Pd-V alloys'', Phys. Rev. B {\bf 52}, 8813 (1994). }
\bibitem{rozen:nature21}{A. Rozen, J. M. Park, U. Zondiner, Y. Cao, D. Rodan-Legrain, T. Taniguchi, K. Watanabe, Y. Oreg, A. Stern, E. Berg, P. Jarillo-Herrero, and S. Ilani, ``Entropic evidence for a Pomeranchuk effect in magic-angle graphene'', Nature {\bf 592}, 214 (2021).}
\bibitem{SM}{See Supplemental Material for (1) derivation of Gibbs entropy for a system of fermions; (2) derivation of the Hamiltonian for the multilayer system.}
\bibitem{won:aip05}{H. Won, S. Haas, D. Parker, S. Telang, A. V\'anyolos, and K. Maki, ``BCS theory of nodal superconductors'', AIP Conf. Proc. {\bf 789}, 3 (2005.)}
\bibitem{alrub:physb08}{T. R. Abu Alrub and S. H. Curnoe, ``Theory of density of states in the nodal superconductor PrOs$_4$Sb$_12$'', Physica B {\bf 403}, 1178 (2008).}
\bibitem{wemple1970polarization}{S. H. Wemple, "Polarization Fluctuations and the Optical-Absorption Edge in BaTiO$_3$", Phys. Rev. B {\bf 2}, 2679 (1970).}
\bibitem{piskunov2004bulk}{S. Piskunov, E. Heifets, R.I. Eglitis, and G. Borstel, "Bulk properties and electronic structure of SrTiO$_3$, BaTiO$_3$, PbTiO$_3$ perovskites: an ab initio HF/DFT study", Comput. Mater. Sci. {\bf 29}, 165 (2004).}
\bibitem{kornich:prr19}{V. Kornich, H. S. Barakov, and Y. V. Nazarov, ``Fine energy splitting of overlapping Andreev bound states in multiterminal superconducting nanostructures'', Phys. Rev. Research {\bf 1}, 033004 (2019).}
\bibitem{kornich:prb20}{V. Kornich, H. S. Barakov, and Y. V. Nazarov, ``Overlapping Andreev states in semiconducting nanowires: Competition of one-dimensional and three-dimensional propagation'', Phys. Rev. B {\bf 101}, 195430 (2020).}
\bibitem{burkov:prl11}{A. A. Burkov and L. Balents, ``Weyl Semimetal in a Topological Insulator Multilayer'', Phys. Rev. Lett. {\bf 107}, 127205 (2011).}
\bibitem{meng:prb17}{T. Meng and L. Balents, ``Weyl superconductors'', Phys. Rev. B {\bf 86}, 054504 (2012).}
\end{thebibliography}
\end{document}